\documentclass[aps,prb,twocolumn,floatfix,showpacs]{revtex4}
\usepackage[dvips]{graphicx}
\usepackage{array}
\usepackage{bm,amssymb,amsmath}
\usepackage{epsf}
\usepackage{psfrag}

\begin{document}

\title{Spatially nonuniform phases in the one-dimensional SU$(n)$ 
Hubbard model \\ for commensurate fillings}

\author{E.~Szirmai, \"O.~Legeza, and J.~S{\'o}lyom}

\affiliation{Research Institute for Solid State Physics and Optics, H-1525
Budapest, P.\ O.\ Box 49, Hungary}

\date{\today}

\begin{abstract}
The one-dimensional repulsive SU$(n)$ Hubbard model is investigated
analytically by bosonization approach and numerically using the density-matrix
renormalization-group (DMRG) method for $n=3,4$, and $5$ for commensurate
fillings $f=p/q$ where $p$ and $q$ are relatively prime. It is shown that the
behavior of the system is drastically different depending on whether $q>n$,
$q=n$, or $q<n$. When $q>n$, the umklapp processes are irrelevant, the model
is equivalent to an $n$-component Luttinger liquid with central charge
$c=n$. When $q=n$, the charge and spin modes are decoupled, the umklapp
processes open a charge gap for finite $U>0$, whereas the spin modes remain
gapless and the central charge $c=n-1$. The translational symmetry is not
broken in the ground state for any $n$. On the other hand, when $q<n$, the
charge and spin modes are coupled, the umklapp processes open gaps in all
excitation branches, and a spatially nonuniform ground state
develops. Bond-ordered dimerized, trimerized or tetramerized phases are found
depending on the filling.
\end{abstract}


\maketitle

\section{Introduction}

Recently, the SU$(n)$-symmetric generalization of the standard SU(2) Hubbard
model\cite{Hubb1-4} has been intensively studied
theoretically.\cite{marston,assaraf,assaraf_half,honer,assaad,szirmai01,szirmai02,buchta_sun}
Apart from its theoretical interest this model may mimic strongly correlated
electron systems where the orbital degrees of freedom of $d$ and $f$ electrons
play important role and these extra degrees of freedom are taken into account
by considering $n$-component fermions. On the other hand, ultracold gases in
optical lattices may also be simulated by such multi-component models.

The Hamiltonian of the model is usually written in the form    
\begin{equation} \begin{split} 
\label{eq:ham}
      {\mathcal H} & =   - t\sum_{i=1}^N\sum_{\sigma=1}^n (c_{i,\sigma}^\dagger 
       c_{i+1,\sigma}^{\phantom \dagger} + c_{i+1,\sigma}^\dagger  
    c_{i,\sigma}^{\phantom\dagger}) \\
       & \phantom{=\,}  +  \frac{U}{2}\sum_{i=1}^N
  \sum_{\substack{\sigma,\sigma'=1 \\ \sigma \neq \sigma'}}^n n_{i, \sigma}n_{i,\sigma'} ,
\end{split}   \end{equation}
where $N$ is the number of sites in the chain. The operator $c_{i,\sigma}^\dagger$
($c_{i,\sigma}^{\phantom \dagger}$) creates (annihilates) an electron at site $i$
with spin $\sigma$, where the spin index is allowed to take $n$ different
values. $n_{i,\sigma}$ is the particle-number operator, $t$ is the hopping integral
between nearest-neighbor sites, and $U$ is the strength of the on-site Coulomb
repulsion.  In what follows $t$ will be taken as the unit of energy.

The model behaves as an $n$-component Tomonaga--Luttinger liquid at generic
fillings. Other type of behavior may appear at commensurate fillings due to
um\-klapp processes. The possible phases, their nature and the critical
coupling where they appear have been studied in detail for two special
commensurate fillings of the band, namely for half filling and $1/n$
filling.\cite{marston,assaraf,assaraf_half,honer,assaad,szirmai01,szirmai02,buchta_sun}
It is well established by now that the ground state is a fully gapped
bond-ordered dimerized state in the half-filled case for any $n>2$. Contrary
to this, the ground state remains translationally invariant in the
$1/n$-filled case, and only the charge mode acquires a gap for $U>U_{\rm
c}$. While Assaraf {\sl et al.} \cite{assaraf} argued that $U_{\rm c}$ is
finite, our recent numerical work\cite{buchta_sun} has suggested a much less, 
perhaps $U_{\rm c}=0$ critical value above which multiparticle umklapp 
processes become relevant.

It is worth mentioning that the SU($n$) Hubbard model has a rich phase diagram
in the attractive case, too.\cite{rapp,zhao} The one-third-filled SU(3) model has
two distinct phases in the high-dimensional limit. In one of them the fer\-mions
form trions while in the other phase a color superfluid state emerges. The
existence of these phases is not yet settled in one dimension.

In the present paper, the role of multiparticle um\-klapp processes will be
further analyzed for general commensurate fillings $f=p/q$, where $p$ and $q$
are relatively prime. We try to establish under what conditions the um\-klapp
processes can generate gaps in the charge or spin sectors, and when and how the
translational symmetry is broken.  To this end, partly analytic, partly
numerical procedures will be applied. We will generalize the method used in
Ref.~[\onlinecite{marston}] to the one-third-filled SU($n$) model to show
analytically that the ground state cannot be spatially uniform. It is
trimerized at least in the large-$n$ limit. The numerical work will show
that in fact this trimerized state with gapped excitations exists for $n > 3$
already.

In the numerical part, the length-dependence of the entropy of finite blocks 
of a long chain is studied. Recently, it has been shown that quantum phase
transitions can be conveniently studied by calculating some measure of
entanglement.\cite{osborne,osterloh,zanardi,gu,vidal,yang,gu2,wu,legeza_qpt} 
This can either be a local quantity, e.g., the concurrence\cite{wootters}, a
global quantity, e.g., the fidelity,\cite{zanardi2} or the entropy of a block of 
several sites.\cite{vidal10} As has been demonstrated recently,\cite{legeza_incomm} 
the oscillatory behavior of the block entropy can reveal the position of soft modes in the 
excitation spectrum of critical systems or the spatial inhomogeneity of gapped
models. This will allow us to demonstrate that at commensurate fillings
$f=p/q$ the type of ground state of the one-dimensional SU($n$) models depends
on whether $q=n$, $q<n$, or $q>n$.

The paper is organized as follows. The oscillatory behavior of the block
entropy, the corresponding peaks in its Fourier spectrum, and their
relationship to the known properties of the half-filled and $1/n$-filled
models are recalled in Sec.~II, where some new results necessitating further
studies are also given. An analytical investigation of the role of umklapp
processes at commensurate fillings is presented in Sec.~III and the
possibility of spatial inhomogeneity of the ground state is discussed. The
numerical results for various fillings are presented in Sec.~IV.  Finally our
findings and conclusions are summarized in Sec.~V.


\section{Oscillatory length dependence of the block entropy}

If a finite block of length $l$ of a long chain of $N$ sites is considered, it
is in a mixed state, even if the long chain is in its ground state. The mixed
state can be described by a density matrix $\rho_N(l)$ and the corresponding 
von Neumann entropy is
\begin{equation}
s_N(l) = -{\text{Tr}}\big[ \rho_N(l) \ln \rho_N(l) \big] \,. 
\label{eq:neumann}
\end{equation} 
It is well known\cite{vidal10,korepin1} that this entropy as a
function of the block size grows logarithmically if the system is critical and
the spectrum is gapless.  In addition, the central charge $c$ can be
derived\cite{holzhey,cardy} from the initial slope of the length dependence 
of $s_N(l)$, 
\begin{equation}   
s_N(l) = \frac{c}{6}\ln \left[ \frac{2N}{\pi} \sin \left( \frac{\pi l}{N}
\right) \right] + g \,,
\label{eq:cardy}
\end{equation}
where $g$ is a shift due to the open boundary. It contains a constant term
which depends on the ground-state degeneracy and an alternating term decaying
with a power of the distance from the boundary.\cite{affleck,laflorence} On
the other hand for noncritical, gapped models, $s_N(l)$ saturates to a finite
value when $l$ is far from the boundaries.

Recently it has been pointed out by some of us \cite{legeza_incomm} that a
wider variety of behavior may be found for the length-dependence of the block
entropy. Namely, we have shown that in some cases oscillations may appear in
$s_N(l)$. This can be best analyzed by considering the Fourier transform
\begin{equation} 
      \tilde{s}(k) = \frac{1}{N}\sum_{l=0}^N e^{- i k l}s_N(l) 
\label{eq:sq}
\end{equation}
for discrete wave numbers, $k=2\pi j/N$ lying in the range $(-\pi,\pi)$. Since
$s_N(l) = s_N(N-l)$, $\tilde{s}(k)$ is real. It has a large peak at $k=0$, all
other Fourier components are negative. Peaks in $|\tilde{s}(k)|$ carry information 
about the position of soft modes or the spatial inhomogeneity of the ground state. 
More precisely if the amplitude of a peak at a nonzero wave number $k^*$ remains
finite in the thermodynamic limit, this indicates a periodic spatial modulation of the 
ground state with wavelength $\lambda=2\pi/k^{\ast}$. On the other hand, if a 
marked peak appears in $|\tilde{s}(k)|$ but its amplitude vanishes as 
$N\rightarrow\infty$, this allows to identify the wave vector of soft modes in critical models.

\begin{figure}[htb]
\includegraphics[scale=0.6]{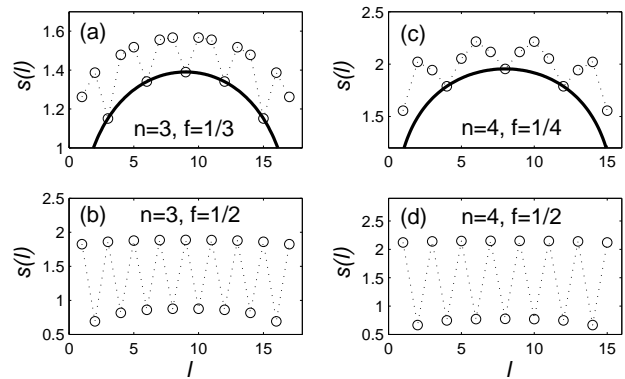}
\caption{Block entropy $s_N(l)$ of finite chains with $N=18$ and $N=16$ site,
respectively, for $n=3$ and 4 at fillings $f=1/n$ and $f=1/2$ for $U=10$. The
solid line is our fit using Eq.~(\ref{eq:cardy}).}
\label{fig:s_l_fig1}
\end{figure}

In a recent work \cite{buchta_sun} we have shown that such oscillations appear
in $s_N(l)$ for the SU($n$) Hubbard model as well. The periodicity depends on
both $n$ and the band filling $f=p/q$. This is shown for the $n=3$ and $n=4$
models for the $1/n$-filled and half-filled cases in Fig.~\ref{fig:s_l_fig1}
for a large value of $U$ $(U=10)$.

In the $1/n$-filled cases, $s_N(l)$ increases logarithmically with the block
length (and then goes down as $l$ approaches $N$). When every third or fourth
values are taken, depending on the periodicity, these selected values can be
fitted to (\ref{eq:cardy}) as shown by the solid lines in panels (a)
and (c).  This indicates gapless behavior and gives $c=n-1$. This is in
agreement with the theoretical expectation, since the charge mode becomes
gapped due to multiparticle umklapp processes and only the $n-1$ spin modes 
are gapless. A distinct behavior is found in half-filled systems, as seen in panels 
(b) and (d). The quantity $s_N(l)$ oscillates with period two, and if only every 
second point is taken, it seems to saturate beyond some block length, before 
decreasing again, indicating that the corresponding models are fully gapped.

\begin{figure}[htb]
\includegraphics[scale=0.5]{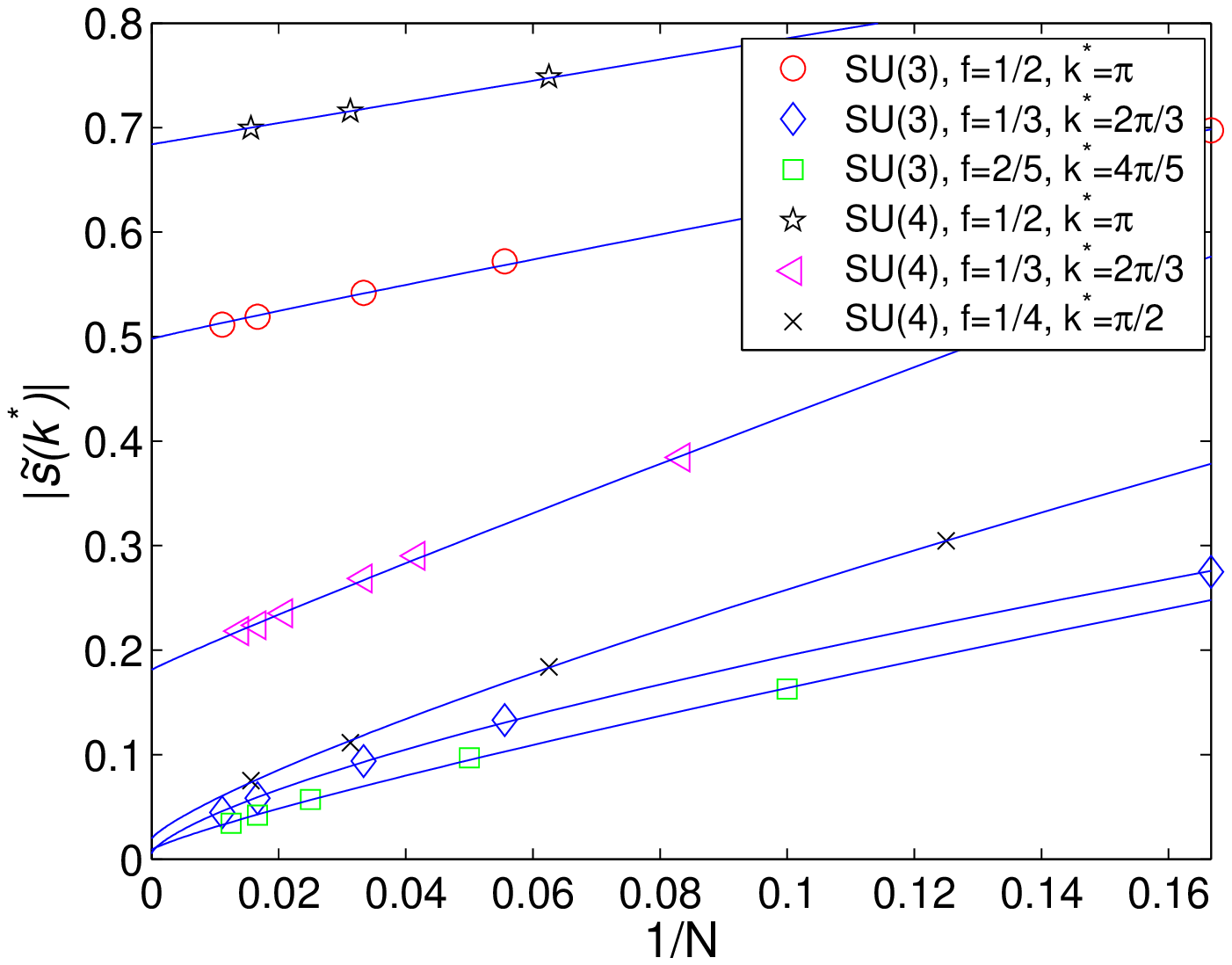}
\caption{Finite-size dependence of $|\tilde s(k^*)|$ for various $n$ and
fillings for $U=10$.  The solid line is the finite-size-scaling fit.}
\label{fig:sq_n}
\end{figure}

The finite-size dependence of the peaks of $|\tilde{s}(k)|$ appearing at 
$k^{\ast}= 2k_{\text{F}} = 2 \pi f$ characterizing
the oscillation is shown in Fig.~\ref{fig:sq_n}. It is seen that in the
$1/n$-filled case the Fourier components at $k^{\ast}=2\pi/n$ vanish in the
thermodynamic limit, while a finite value is obtained at $k^{\ast}=\pi$ for
half-filled models.  This corroborates our finding that the $1/n$-filled
SU($n$) models are critical with a spatially uniform ground state while a
gapped bond-ordered dimerized phase appears at half filling.

\begin{figure}[htb] 
\includegraphics[scale=0.6]{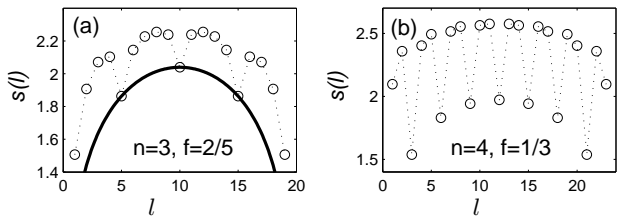} 
\vspace{-1mm}\hspace{6cm} $l$ 
\hspace{4cm} $l$ \\  
\vspace{-3mm}
\caption{Same as Fig.~\ref{fig:s_l_fig1} but for (a) $n=3, f=2/5$, $N=20$ and (b)
$n=4, f=1/3$, $N=24$.}
\label{fig:s_l_fig2} 
\end{figure}

We have done similar calculations for more general commensurate fillings of
the band. Figure~\ref{fig:s_l_fig2} shows the results obtained for
$n=3$, $f=2/5$ as well as for $n=4$, $f=1/3$. When every fifth points
are taken for the $f=2/5$ filled SU(3) model, they can be fitted to
\eqref{eq:cardy} yielding $c=3$ for the central charge. This indicates that
all modes, including the charge mode, are gapless. 

Such a fit does not work for the one-third-filled SU(4) model. To better see the 
difference $|\tilde{s}(k)|$ is considered again. The amplitude of the Fourier 
component at $k^{\ast}= 4\pi/5$, also displayed in Fig.~\ref{fig:sq_n} for the $n=3$ 
model, vanishes in the $N\rightarrow\infty$ limit. On the other hand $|\tilde{s}(k)|$ 
remains finite at $k^{\ast} = 2\pi/3$ in the one-third-filled $n=4$ model.

\begin{figure}[htb]
\includegraphics[scale=0.5]{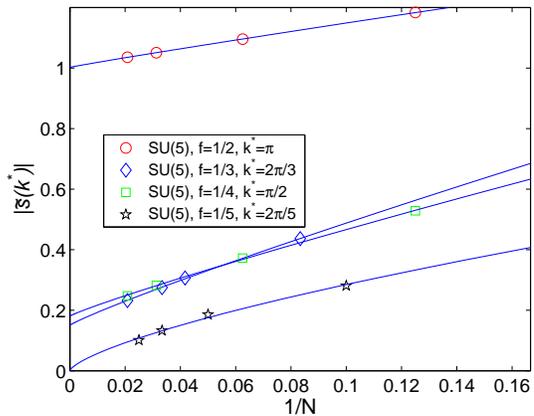}
\caption{Same as Fig.~\ref{fig:sq_n} but for $n=5$.
The solid line is the finite-size-scaling fit.}
\label{fig:su5_sq_n}
\end{figure}

When the same calculations are repeated for the $n=5$ model at $f=1/2, 1/3,
1/4$, and $1/5$, peaks appear in $|\tilde s(k)|$ at $k^*=\pi, 2\pi/3$, $\pi/2$,
and $2\pi/5$, respectively.  As is seen in Fig.~\ref{fig:su5_sq_n}, the
amplitude of the peaks remains finite even when $N\rightarrow\infty$ in the
first three cases, while it vanishes in the last case.

These results indicate that the role of umklapp processes depends on the
relationship between the number of components $n$ and the relative primes $p$
and $q$ characterizing the commensurate filling. In what follows this problem
will be studied first analytically in a bosonization approach and large-$n$
expansion technique, and then numerically using the DMRG method.


\section{Analytical considerations}

\subsection{The role of umklapp processes: a bosonization approach}

Following the usual procedure we write the Hamiltonian \eqref{eq:ham} in
momentum space and linearize the free-particle spectrum around the two Fermi
points $(\pm k_{\text{F}})$. The underlying assumption is that the low-lying
excitations determine the physics of the system. Depending on whether the
momentum of the fermions is close to $+k_{\text{F}}$ or $-k_{\text{F}}$, one
can distinguish left- and right-moving particles, and the interaction
processes also can be classified on the basis of whether the incoming and
scattered particles are right or left movers and the momentum transfer is
small (forward scattering) or large, of the order of $2k_{\text{F}}$ (backward
scattering). In a generic model the strength of the various scattering
processes may be different. For the sake of simplicity we neglect chiral
processes in which both particles move in the same direction before and after
the interaction, since they lead to the renormalization of the
Fermi velocity only.

One can recognize that at generic fillings, where um\-klapp processes do not
play a role, the forward and backward scattering processes can be interpreted
as current--current interactions and their contribution to the Hamiltonian
density can be conveniently rewritten using Dirac fermions\cite{zinn-justin}
in the following short form (automatic summation for the repeated indices is
understood):
\begin{equation}
\label{eq:intham}
H_{\textrm{int}}(x) = {\textstyle{\frac{1}{2}}}
g_{\sigma_1 \sigma_2 \sigma_3 \sigma_4}\bar\psi_{\sigma_1}(x)\gamma_\mu
    \psi_{\sigma_2}(x)\bar\psi_{\sigma_3}(x)\gamma_\mu\psi_{\sigma_4}(x) .
\end{equation}
Here $\sigma_i$ denote the spin indices that can take the values $1,\ldots, n$, 
$\gamma_\mu$ with $\mu=1,2$ are the Dirac matrices, in our case the standard 
Pauli matrices ($\sigma_x$, $\sigma_y$), and $\bar\psi(x)=\psi^\dagger(x)\gamma_1$. 
While the Hubbard model contains a single interaction parameter $U$, the couplings 
$g_{\sigma_1 \sigma_2 \sigma_3 \sigma_4}$ may be different for physically different 
processes in more realistic models. In the renormalization-group treatment we will 
assume this to be the case. It is assumed, however, that the spin of the fermions does 
not change in the scattering process and the couplings are symmetric under the exchange 
$(\sigma_1, \sigma_2) \leftrightarrow (\sigma_3, \sigma_4)$. If the fermion field 
$\psi_{\sigma}(x)$ is decomposed into left- and right-moving components according to
\begin{equation} 
\psi_{\sigma}(x)=\begin{pmatrix} 
R_{\sigma}(x) \\ L_{\sigma}(x) 
\end{pmatrix}, \hskip 0.4cm  
\hskip 0.4cm 
  \bar\psi_{\sigma}(x)=\left( L^\dagger_{\sigma}(x), \,  R^\dagger_{\sigma}(x) \right), 
\end{equation}  
the usual backward- and forward-scattering processes are in fact recovered. 
In the standard $g$-ology\cite{solyom} notation $g_{\sigma \sigma' \sigma' \sigma}$ is denoted 
by $-g_1$, and $g_{\sigma \sigma \sigma' \sigma'}$ by $g_2$. 

The well-known renormalization-group (RG) equations, the $\beta$ function can
be written for these scattering processes in a short form:\cite{zinn-justin} 
\begin{equation}   \begin{split}
\frac{\partial \ln g_{\sigma_1 \sigma_2 \sigma_3 \sigma_4}}{\partial \ln \Lambda'/\Lambda} \equiv 
    \beta_{\sigma_1 \sigma_2 \sigma_3 \sigma_4 } & =
g_{\sigma_1 \sigma_i \sigma_3 \sigma_j}g_{ \sigma_i \sigma_2 \sigma_j \sigma_4} \\
     & \phantom{=\, }- 
   g_{\sigma_1 \sigma_i \sigma_j \sigma_4}g_{\sigma_i \sigma_2 \sigma_3 \sigma_j},
\end{split}    \end{equation}
where $\Lambda$ is the cut-off parameter. These RG equations have been analyzed 
earlier\cite{marston,szirmai01} and it was found that the backward-scattering processes 
scale out at generic fillings
in the SU($n$) Hubbard model and for this reason this model is equivalent to
an $n$-component Luttinger liquid in this case. The Hamiltonian can be
diagonalized\cite{szirmai02} and the excitation spectrum can be determined
exactly in bosonic phase-field representation.\cite{bozonizacio} There is one symmetric
combination of the phase fields with different spin indices, this is the
so-called charge mode:
\begin{equation}
\label{eq:modec}
\phi_\mathrm{c}(x)  =  \frac{1}{\sqrt{n}}\sum_{\sigma=1}^n\phi_{\sigma}(x), 
\end{equation}
while the $n-1$ antisymmetric combinations give the spin modes:
\begin{equation}
\label{eq:modes}
\phi_{m\mathrm{s}}(x)  = \frac{1}{\sqrt{m(m+1)}}\left[ \sum_{\sigma=1}^m
\phi_{\sigma}(x) - m\phi_{m+1}(x) \right]  
\end{equation}
with $m=1,\ldots ,n-1$. 

Similarly to the spin-charge separation in the two-component Luttinger model,
one finds complete mode separation. The Hamiltonian density is the sum 
of the contributions of the individual modes,
\begin{equation}
    H(x) = \sum_j H_j(x) \,,
\end{equation}
where $j={\text{c}},1{\text{s}},2{\text{s}},\ldots ,(n-1){\text{s}}$. Each term
has the usual bosonic form:
\begin{equation}
\label{eq:newham}
H_j(x) = \frac{\hbar u_j}{2}\left\{  \pi K_j \left[\Pi_j(x)\right]^2 
+ \frac{1}{\pi K_j} \left[\partial_x\phi_j(x)\right]^2 \right\} ,
\end{equation}
where $\Pi_j(x)$ is the momentum canonically conjugated to $\phi_j(x)$. The
renormalized velocities and the Luttinger parameters can be given in terms of
the new couplings $g_{2;j}$ appearing after diagonalization\cite{szirmai02} 
in the spin indices:
\begin{subequations}
\begin{flalign}
u_j = & \, v_\textrm{F}\sqrt{1-g_{2;j}^2}\,, \\ K_j = & \,
\sqrt{\frac{1-g_{2;j}}{1+g_{2;j}}} \,.
\end{flalign}
\end{subequations}

In a finite system, where the momentum is quantized in units of $2\pi/L$, the
excitation spectrum of the Luttinger model can be written as:
\begin{equation}
\label{eq:Lutt-Ham-1}
E = \sum_j \hbar u_j \frac{2\pi}{L} \left( n_+^j + n_-^j + \Delta_+^j +
\Delta_-^j \right),
\end{equation}
where $n_\pm ^j$ are integers describing the bosonic excitations and 
\begin{equation}
\label{eq:Lutt-Ham-2}
    \Delta_\pm^j = \frac{1}{16} \Big( \sqrt{K_j}  J_j \pm
    \frac{1}{\sqrt{K_j}} \delta N_j \Big)^2  \,,
\end{equation}
where $\delta N_j$ is the change in the number of particles in the $j$th
channel, and similarly $J_j$ describes the current in the $j$th channel
created by adding particles to or removing them from the branches of the
dispersion relation.

The total momentum is given by 
\begin{equation} 
P=\hbar k_\textrm{F} J_c + \sum_j \hbar \frac{2 \pi}{L}\left(n_+^j -
n_-^j + \Delta_+^j - \Delta_-^j \right).
\end{equation}
Thus soft modes appear not only at zero momentum but at integer multiples 
of $2k_\textrm{F}$, too, since the charge current $J_c$ is an even number if the 
total charge is conserved. Since
\begin{equation}
2 k_\textrm{F}=\frac{N_{\text{c}}^0}{n}\frac{2 \pi}{L} \,, 
\end{equation}
where $N_{\text{c}}^0$ is the number of particles in the system, and the
filling of the band is $f = N_{\text{c}}^0/nN$, the position of these soft
modes depends on the filling only, $2k_{\text{F}}= 2 \pi f$.

\begin{figure}[htb] 
\psfrag{kf=1/2}{$k_\textrm{F}=\pi /2$}
\psfrag{kf=1/3}{$k_\textrm{F}=\pi /3$}
\psfrag{n=3}{$n=3$}
\includegraphics[scale=0.38]{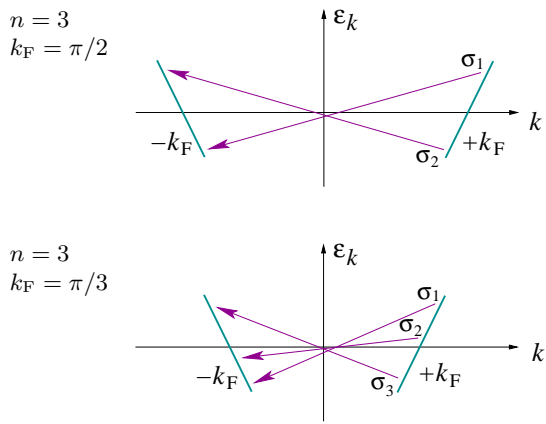}
\caption{Umklapp processes in the SU(3) symmetric Hubbard model. (a) Any two
of the three types of fermion can be scattered in the half-filled case. (b) Fermions with 
the three possible spin orientations participate in the scattering process in the 
one-third-filled case.}
\label{fig:umklappok}
\end{figure}

We know that the usual umklapp processes, scattering of two right movers into
left-moving states or vice versa, that were neglected so far, are relevant in
a half-filled system for any $n$.\cite{marston,szirmai01} Multiparticle umklapp
processes may become relevant at other commensurate fillings. To see what kind
of processes are allowed we have to take into account that the total
quasimomentum transferred in an umklapp process has to be an integer multiple
of $2\pi$. If the band is $f=p/q$ filled and consequently $k_{\text{F}} = p\pi/q$, 
$q$ particles have to be scattered from one Fermi point to the opposite
one to satisfy this condition.  This is illustrated in
Fig.~\ref{fig:umklappok}. Since there are $n$ different types of fermion in
the SU($n$) model, and the Hubbard interaction is local, only such umklapp
processes are allowed by the Pauli principle in which the number of scattered
particles is less than or equal to $n$. Thus $q$-particle umklapp processes
are forbidden in our model when $q > n$.
  
The role of multiparticle umklapp processes in the conductivity of the
SU(2) Hubbard model has been studied in Ref.~[\onlinecite{giamarchi}].  A
different aspect, whether the charge and spin modes are coupled or not by the
possible multiparticle umklapp processes has been considered for the SU($n$)
model.\cite{buchta_sun} The $q$-particle umklapp processes can be described in
the bosonic phase-field representation by
\begin{equation}
\label{eq:umkl}
H_\textrm{U} (x) = g_3 \int \textrm{d}x \sum_{\{\sigma_i\}'}
\cos\big[2\left(\phi_{\sigma_1}(x)+\ldots +\phi_{\sigma_q}(x)\right)\big] \,.
\end{equation}
The phase fields appearing here are the phases of the bosonic representation
of the particles participating in the scattering process, and $\{\sigma_i\}'$
indicates that all spin indices have to be different.

This Hamiltonian can be expressed in terms of the phase fields corresponding
to the charge and spin modes, making use of the inverse of (\ref{eq:modec}) and
(\ref{eq:modes}). It is clear that if $q=n$, only the symmetric combination of
the phase fields, that is the charge mode defined in (\ref{eq:modec}) appears
in \eqref{eq:umkl}. This means that in this case the $q$-particle umklapp
processes couple to the charge mode only. The Hamiltonian density of the
charge mode is identical to that of the well-known sine-Gordon model which has
a fully gapped excitation spectrum. Thus the charge mode becomes gapped 
for finite $U$. The $n-1$ spin modes are not influenced by the umklapp 
processes in the $1/n$-filled case. They remain gapless and the central
charge is $c=n-1$.

When $q<n$, the sum of the $q$ phase fields corresponding to the $q$ particles
required for umklapp processes, will contain various combinations of the $n$
boson fields, leading to a mixing (coupling) of the charge and spin modes,
thus opening gaps in all modes. The model becomes noncritical for $U >
U_{\text{c}}$.
 
When on the other hand $q>n$, umklapp processes are forbidden by 
Pauli's exclusion principle when the interaction is local.  The charge and 
spin modes remain gapless, thus $c=n$ for $U>0$. The expected behavior for 
different cases is summarized in Table~\ref{tab:analitikus}.

\begin{table}[htb]
\begin{tabular}{@{}ccccccccc@{}}  \hline \hline 
                  & \phantom{+} & $n$ & \phantom{+}& $c$ & \phantom{+} & phase
        & \phantom{+} & $k^*$ \\ \hline $q=n$ & & any $n$ & & $n-1$ & &
        C0S$(n-1)$ & & $2\pi p/n$ \\ $q<n$ & & $n\neq 2$ & & -- & & C0S0 & &
        $2\pi p/q$ \\ $q>n$ & & any $n$ & & $n$ & & C1S$(n-1)$ & & $2\pi p/q$
        \\ \hline \hline
\end{tabular}
\caption{Central charge $c$ and the type of phase characterized by the 
number of soft modes in the charge and spin sectors (CxSy) for the
$p/q$-filled SU($n$) Hubbard model.} 
\label{tab:analitikus}
\end{table}

\subsection{Spatial inhomogeneity in the large-$n$ limit 
of the SU($n$) Hubbard-Heisenberg model at one-third filling}

We have seen that the spectrum is fully gapped when $q<n$. The question
naturally arises whether the opening of a gap at multiples of
$k^*=2k_{\text{F}}$ is related to a breaking of the translational symmetry, an
instability against the formation of a spatially inhomogeneous state with the
corresponding wave number.

To analyze the stability of the homogeneous state we generalize the procedure used 
by Marston and Affleck\cite{marston} to the one-third-filled case in
the large-$U$ limit. When $n$ is an integer multiple of 3, the number of
fermions sitting on each site is an integer in a homogeneous sample, and a
finite energy is needed to add an extra particle. This energy gap at
$k_{\text{F}}$ may imply a tripling of the spatial period in the ground state.
To search for this spatial inhomogeneity a more general model, the SU($n$)
symmetric generalization of the Hubbard-Heisenberg model will be considerd.
Its Hamiltonian is
\begin{eqnarray}
\label{eq:hub-hei-ham} 
H &=&\sum_{i=1}^N \bigg[ -  \frac{J}{n} \sum_{\sigma,\sigma'=1}^n (c_{i,\sigma}^\dagger 
    c^{\phantom \dagger}_{i+1,\sigma})(c_{i+1,\sigma'}^\dagger c^{\phantom \dagger}_{i,\sigma'}) \\ 
      &- &  t\sum_{\sigma=1}^n (c_{i,\sigma}^\dagger c^{\phantom \dagger}_{i+1,\sigma} + 
      \text{h.c.} ) +  \frac{U}{n} \bigg( \sum_{\sigma=1}^n c_{i,\sigma}^\dagger 
    c^{\phantom \dagger}_{i,\sigma} - \frac{n}{3} \bigg)^2 \bigg].  \nonumber
\end{eqnarray}
The chemical potential is shifted to zero at one-third filling, and the Hubbard coupling $U$ 
and Heisenberg coupling $J$ are rescaled by $2/n$
so that the spacing of the energy levels remain the same as $n$ increases. We
note that the usual $J\sum_{<i,j>}\mathbf{S}_i\mathbf{S}_j$ nearest-neighbor
Heisenberg interaction breaks up into three terms in fermionic representation
of the spin operators. One of them only shifts the chemical potential, another 
corresponds to nearest-neighbor Coulomb interaction (that is
unimportant in the large $U$ limit), therefore they are neglected.

The equilibrium state will be determined from the minimum of the free energy
that can be derived from the partition function of the system.  In
functional integral formalism the partition function can be expressed with the
Lagrangian of the model. At finite temperatures the imaginary time Lagrangian
is $L[c,c^\dagger]=\sum_{i,\sigma} \big(c_{i,\sigma}^\dagger (\textrm{d}/\textrm{d}\tau)
c^{\phantom \dagger}_{i,\sigma} + H\big)$ and the partition function is 
\begin{equation}
\label{eq:Z}
Z=\int [\textrm{d}c][\textrm{d}c^\dagger ]\textrm{exp}\bigg( -\int_0^\beta
\textrm{d}\tau \, L[c,c^\dagger] \bigg).
\end{equation}
Here $\beta$ is the inverse temperature. The integral occurring in the above
expression cannot be calculated in a simple way due to the quartic terms in
the Lagrangian.  However, these quartic terms can be eliminated by a 
Hubbard-Stratonovich transformation based on the integral identity
\begin{equation}
\label{eq:hub-strat}
\textrm{exp}(VX^2) \propto \int \textrm{d}Y \textrm{exp} (-Y^2/4V + XY).
\end{equation}
In our case the quantities corresponding to $X$ are $\sum_{\sigma} 
c_{i,\sigma}^\dagger c^{\phantom \dagger}_{i,\sigma}$ and 
$\sum_{\sigma} c_{i,\sigma}^\dagger c^{\phantom \dagger}_{i+1,\sigma}$. 
Therefore we have to introduce $2N$ bosonic fields: $\phi_i$ and $\chi_{i,i+1}$.
We note that the fields $\phi_i$ and $\chi_{i,i+1}$ correspond\cite{marston} to
site- and bond-centered densities, respectively. 

Adding the appropriate terms to the Lagrangian gives
\begin{multline} 
\label{eq:lagr-def}
L[c,c^\dagger,\phi,\chi] \stackrel{\textrm{def}}{=} L[c,c^\dagger] \\ 
   + \frac{n}{U} \sum_i \bigg[ \frac{\phi_i}2 + i\frac{U}{n} \Big(\sum_{\sigma}
c_{i,\sigma}^\dagger c^{\phantom \dagger}_{i,\sigma} - \frac{n}{3}\Big) \bigg]^2 \\ +
\frac{n}{J} \sum_i \bigg| \frac{1}{2}\chi_{i,i+1} + \frac{J}{n} \sum_{\sigma}
c_{i,\sigma}^\dagger c^{\phantom \dagger}_{i+1,\sigma} \bigg|^2.
\end{multline}
The new Lagrangian is quadratic in the fields, however the fermionic fields are 
coupled to the bosonic ones. The explicit form of the Lagrangian is 
\begin{multline}
\label{eq:lagr}
L [ c,c^\dagger ,\phi ,\chi ] = \sum_{i,\sigma} \bigg\{ c_{i,\sigma}^\dagger
\Big(\frac{\textrm{d}}{\textrm{d}\tau} + i \phi_i \Big) c^{\phantom \dagger}_{i,\sigma}
+\frac{1}{4J} | \chi_{i,i+1} |^2 \\ 
   + \big[ ( \chi_{i,i+1} - t ) c_{i,\sigma}^\dagger c^{\phantom \dagger}_{i+1,\sigma} + h.c. \big] 
    + \frac{1}{4U}  \phi_i^2 - i \frac{1}{3} \phi_i \bigg\}.
\end{multline}
The partition function is obtained by integrating over $c$ and $c^\dagger$:
\begin{multline}
Z[\phi,\chi] = \int [\textrm{d}c][\textrm{d}c^\dagger ]\textrm{exp}\bigg(
-\int_0^\beta \textrm{d}\tau \, L[c,c^\dagger,\phi,\chi] \bigg) \,.
\end{multline}
Writing it in the form $Z[\phi,\chi]\equiv
\textrm{exp}(-S_{\textrm{eff}}[\phi,\chi])$, this defines the effective
action. The free energy can be expressed in a usual way via $Z[\phi,\chi]$ as
\begin{equation}
F[\phi,\chi]=-1/\beta \, \, \textrm{ln}\big( Z[\phi,\chi] \big).
\end{equation}

As mentioned in the previous subsection, if spatial oscillations occur
in the system, they are expected to appear with wave number $k^*=2\pi f$.
Thus we may expect spontaneous trimerization at one-third filling. Therefore 
we suppose that the boson field $\phi_i$ takes three
different values depending on whether $i=3l$, $i=3l+1$ or $i=3l+2$ with 
integer $l$.  They will be denoted as $\phi_1$, $\phi_2$, and $\phi_3$. 
Similar assumption holds for the fields
$\chi_{i,i+1}$, too. The three values are $\chi_1$, $\chi_2$, and $\chi_3$. The
lattice is thus decomposed into three sublattices.

To get a real free energy the fields $\phi_\alpha$ ($\alpha=1,2,3$) are
redefined by continuing to the complex plane ($i\phi_\alpha\rightarrow
\phi_\alpha$). The free energy can then be written as
\begin{multline}
\label{eq:free-en}
F(\{\phi_\alpha \},\{\chi_\alpha \})= \frac13 \sum_{\alpha=1}^3 \left(
\frac{Nn}{4J}\chi_\alpha^2 + \frac{Nn}{4U} \phi_\alpha^2 -
\frac{Nn}{3}\phi_\alpha \right) \\ + \frac n3 \sum_k \Big( E(k) - 1/3 \Big),
\end{multline}
and the summation for $k$ has to be performed over the
reduced Brillouin zone which is now one-third of the original one ($k$ runs
from $-\pi/3$ to $\pi/3$) and $E(k)$ is the energy spectrum of a single
fermion coupled to the boson fields. It is the eigenvalues of the Hamiltonian
\begin{multline}
\label{eq:sp-Ham}
H = \sum_i\Big[ \Big( (\chi_1-t)a_i^\dagger b^{\phantom \dagger}_{i+1} +
(\chi_2-t)b_{i+1}^\dagger c^{\phantom \dagger}_{i+2} \\ + (\chi_3-t)c_{i+2}^\dagger 
   a^{\phantom \dagger}_{i+3} + {\text{h.c.}} \Big) \\
     + \phi_1 a_i^\dagger a^{\phantom \dagger}_i + \phi_2 b_{i+1}^\dagger
b^{\phantom \dagger}_{i+1} + \phi_3 c_{i+2}^\dagger c^{\phantom \dagger}_{i+2} \Big] ,
\end{multline}
where the operators $a$, $b$ and $c$ belong to different sublattices. In order to 
determine the one-particle spectrum one has to diagonalize Hamiltonian
\eqref{eq:sp-Ham} in momentum space. Therefore, we are looking for the
eigenvalues of the matrix
\begin{equation}
\label{eq:matrix}
\begin{pmatrix}
\phi_1 & (\chi_1 - t){\text{e}}^{-ik} & (\chi_3 - t){\text{e}}^{ik} \\ 
(\chi_1 - t){\text{e}}^{ik} & \phi_2 & (\chi_2 - t){\text{e}}^{-ik} \\
(\chi_3 - t){\text{e}}^{-ik} & (\chi_2 - t){\text{e}}^{ik} & \phi_3
\end{pmatrix}.
\end{equation}
The energy spectrum $E(k)$ has three branches corresponding to the three
(real) solutions of the third-order eigenvalue equation:
\begin{subequations}
\begin{flalign}
\label{eq:spectrum}
 & E_1(k) - (\phi_1 + \phi_2 + \phi_3)/3 =\\
    & \phantom{E_1(k)}  -\textrm{sign}(Q)\sqrt{|P|}
  \cos\bigg[ \frac13 \cos^{-1} \bigg(\bigg|\frac{Q}{(|P|)^{3/2}}\bigg|\bigg)
  \bigg]  ,  \nonumber   \\
 & E_{2,3}(k)  -  (\phi_1 + \phi_2 + \phi_3)/3 =   \\
  &  \phantom{\,\,} -\textrm{sign}(Q)\sqrt{|P|} \cos\bigg[ \frac13 \cos^{-1}
  \bigg(\bigg|\frac{Q}{(|P|)^{3/2}}\bigg|\pm \frac{2\pi}{3}\bigg) \bigg] .   \nonumber
\end{flalign}
\end{subequations}
Here $P$ and $Q$ are the parameters in the eigenvalue equation when
transformed to the form $\tilde{E}^3(k) + P \tilde{E}(k) + Q =0$ with 
$\tilde{E}(k)=E(k)-(\phi_1 + \phi_2 + \phi_3)/3$. 

The minimum of the free energy of a system with such a spectrum cannot be evaluated
quite generally. Fortunately, we are only interested whether the three $\phi_{\alpha}$
and $\chi_{\alpha}$ are different or not. This analysis can be carried out easier in terms
of the linear combinations:
\begin{subequations}
\begin{flalign}
\phi := & \, (\phi_1 + \phi_2 + \phi_3)/3,\\
\Delta\phi_{1} := & \, (\phi_1 - \phi_2 )/2,\\ 
\Delta\phi_{2} := & \, (\phi_1 +\phi_2 - 2\phi_3)/6,
\end{flalign}
\end{subequations}
and similar definitions for $\chi$, $\Delta\chi_1$ and $\Delta\chi_2$. One finds that 
although the free energy has an extremum at $\Delta\phi_{1}=\Delta\phi_{2}=0$ 
and $\Delta\chi_{1}=\Delta\chi_{2}=0$, the free energy of the uniform state is not 
a local minimum. Thus a density-wave has to appear in the system. We can conclude 
that the SU($n$) Hubbard model is unstable against the Heisenberg coupling in the 
large-$n$ limit and exhibits inhomogeneous spatial ordering for $U>0$. 

>From this analysis alone -- due to the rather complicated one-particle spectrum -- we 
cannot decide whether the system is dimerized, trimerized, or some other periodicity 
occurs, and whether the density wave is site-centered or bond-centered. It is natural to 
relate the nonuniform phase to the fully gaped excitation spectrum. Thus when our 
previous considerations are taken into account, trimerized phase is expected in 
one-third-filled models. This will be supported by the numerical calculation. We will also 
see, that the trimerized phase is not a special feature of the systems with integer number 
of electrons per site. It occurs in one-third-filled system for arbitrary $n>3$. 


\section{Numerical study of the spatial inhomogeneity}

In this section we present our numerical results obtained by the
DMRG\cite{white} method for the length dependence of the block entropy
$s_N(l)$ and its Fourier transform $\tilde{s}(k)$ to relate them to
the number and position of soft modes when the model is critical, or to
spatial inhomogeneity of the ground state for gapped models.

The spatial modulation of the ground state can be a site- or a bond-centered
density wave. A site-centered density wave would manifest itself in an
oscillation of the von Neumann entropy of single sites, $s_i$, with
$i=1,\dots, N$ or in the local electron density defined by
\begin{equation}
\langle n_{i} \rangle = \sum_{a=1}^n \langle \Psi_{\rm GS}| n_{i,a} |
\Psi_{\rm GS} \rangle\,,
\label{eq:n_i}
\end{equation}
where $|\Psi_{\rm GS} \rangle$ is the ground-state wavefunction. The wave
number of the charge oscillation can again be determined from peaks in the
Fourier transform of $s_i$ or $\langle n_i \rangle$ denoted as $s_1(k)$ and
$n(k)$, respectively.

The existence of a bond-centered density wave can be demonstrated by
studying the variation of the bond energy or the two-site entropy along the chain.
To avoid boundary effects we have calculated the difference of two-site 
entropies in the middle of the chain, between first, second, third and so on 
neighbor bonds:
\begin{subequations}
\begin{flalign}
   D_s(N)  = & s_{N/2,N/2+1}-s_{N/2+1,N/2+2} \,, 
\label{eq:d_s}    \\
     T_s(N)   = & s_{N/2,N/2+1}-s_{N/2+2,N/2+3} \,, 
\label{eq:t_s}  \\
     Q_s(N)  = & s_{N/2,N/2+1}-s_{N/2+3,N/2+4} \,, 
\label{eq:q_s}   \\
     P_s(N)   = & s_{N/2,N/2+1}-s_{N/2+4,N/2+5} \,.
\label{eq:p_s}
\end{flalign}
\end{subequations}
For convenience the number of sites in the chain was always even. 
Moreover, since we expect dimerized, trimerized or tetramerized phases
depending on the commensurate filling $p/q$, the number of sites $N$ was 
always taken to be an integer multiple of $q$.

When a doubling of the lattice periodicity of the ground state is indicated by 
a finite peak in $|\tilde s(k)|$ at $k^*=\pi$, a truly dimerized phase gives 
equal finite values for $D_s$ and $Q_s$ and vanishing $T_s$ and $P_s$ 
in the $N\rightarrow\infty$ limit. Stronger and a weaker bonds alternate 
along the chain. When the peak in $|\tilde s(k)|$ appears at $k^*=2\pi/3$ 
and a trimerized phase is expected, $T_s$ should be finite and $Q_s$
should vanish. Symmetry considerations imply that two equally strong bonds 
are followed by a weaker or stronger bond in this case. In a tetramerized 
phase, the peak in $|\tilde s(k)|$ appears at $k^*=\pi/2$, $D_s$, $T_s$, 
and $Q_s$ may be finite in the $N\rightarrow\infty$ limit and only $P_s$ 
vanishes necessarily. 

\subsection{The numerical procedure}

The numerical calculations presented in this paper have been performed on
finite chains with open boundary condition (OBC) using the DMRG
technique and the dynamic block-state selection (DBSS)
approach.\cite{legeza_dbss,legeza_qdc} We have set the threshold value of the
quantum information loss $\chi$ to $10^{-5}$ for $n=3,4$ and to $10^{-4}$ for 
$n=5$ and the minimum number of block states $M_{\rm min}$ to $256$. 
In spite of the large number of degrees of freedom per site in the $n=5$ case the 
entropy analysis allows one to study this problem as well. The ground state has been 
targeted using four to eight DMRG sweeps until the entropy sum rule has been
satisfied.  The accuracy of the Davidson diagonalization routine has been set
to $10^{-7}$ and the largest dimension of the superblock Hamiltonian was
around three millions. As an indication of the computational resources used in the present work 
we note that the maximum number of block states was around 1600 for $n=3$ and 
900 for $n=4$ and $5$. 

The large-$N$ limit of the entropies and amplitudes of the peaks in the
Fourier spectrum can be obtained if appropriate scaling functions are used. In
a critical, gapless model, in leading order, these are expected to scale to
zero as $1/N$ while in a noncritical model the scaling function depends on the boundary
condition.  Therefore for any quantity $A$ the finite-size scaling ansatz
\begin{equation}
    A(N) = A_0 + a/{N^\beta}
    \label{eq:scale_obc}
\end{equation}
is used to evaluate the data obtained with OBC, where $A_0$, $a$, $\beta$ are
free parameters to be determined by the fit.


\subsection{The numerical results}

\subsubsection{Models with $q=n$}

The $1/n$-filled case ($q=n$) has already been considered in
Ref.~\onlinecite{buchta_sun} and some of the results were listed in
Sec.~II. As has been shown in Fig.~\ref{fig:s_l_fig1}, $s_N(l)$ oscillates
with period $n$ for finite systems. These oscillations are due to the soft modes located 
at wave numbers $k^{\ast} = 2\pi/n$. Taking every $n$th value only, $s_N(l)$ can
be fitted accurately using (\ref{eq:cardy}) for relatively short chains already if $U$ is large.
After a proper finite-size scaling, the fit gives $c=n-1$, as expected.

\begin{figure}[htb]
\includegraphics[scale=0.5]{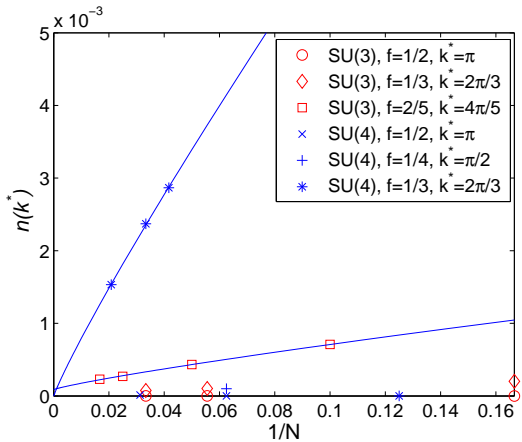}
\caption{Finite-size dependence of $n(k^*)$ for various $n$ and fillings for
$U=10$.  The solid line is the finite-size-scaling fit.}
\label{fig:nq_n}
\end{figure}

Taking the Fourier transform of $s_N(l)$, besides the large positive peak at $k=0$,
additional negative peaks are found at the positions of the soft modes, at $k^*=\pi$, 
$2\pi/3$, $\pi/4$, and $2\pi/5$ for $n=2,3,4,5$, respectively. Their amplitude vanishes, 
however, in the $N \rightarrow \infty$ limit. As a further check that there are neither 
site- nor  bond-centered oscillations in the ground state we have analyzed $s_i$, 
$\langle n_i \rangle$ and $s_{i,i+1}$. All Fourier components of these quantities scale 
to zero in the thermodynamic limit. As an example, the finite-size dependence of
$n(k^*)$ for $n=3$ and $n=4$ is shown in Fig.~\ref{fig:nq_n} at $U=10$.

\begin{figure}[htb]
\includegraphics[scale=0.5]{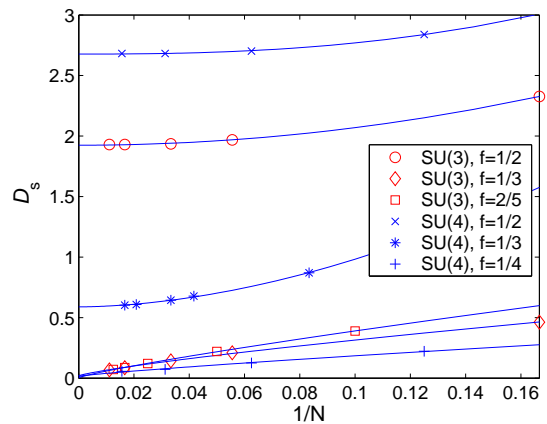}
\caption{Finite-size dependence of $D_s$ for various $n$ and fillings
for $U=10$.  The solid line is the finite-size-scaling fit.}
\label{fig:ds_n}
\end{figure}

\begin{figure}[htb]
\includegraphics[scale=0.5]{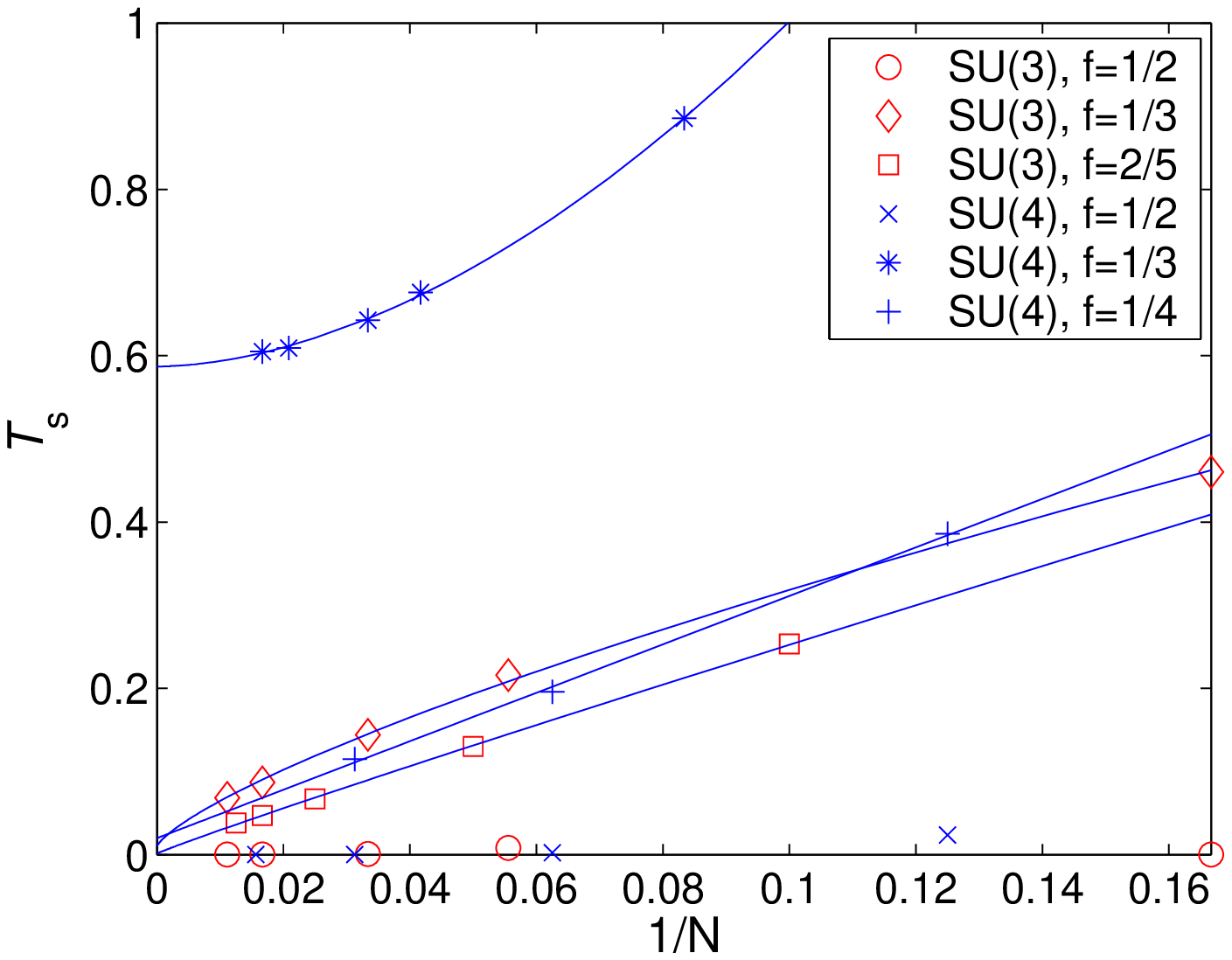}
\caption{Finite-size dependence of $T_s$ for various $n$ and fillings
for $U=10$.  The solid line is the finite-size-scaling fit.}
\label{fig:ts_n}
\end{figure}

\begin{figure}[htb]
\includegraphics[scale=0.5]{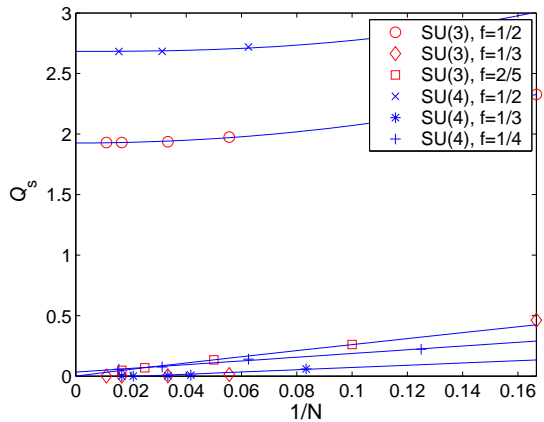}
\caption{Finite-size dependence of $Q_s$ for various $n$ and fillings
for $U=10$.  The solid line is the finite-size-scaling fit.}
\label{fig:qs_n}
\end{figure}

When the finite-size scaling of $D_s$, $T_s$, and $Q_s$ is
analyzed one finds that they all vanish in the thermodynamic limit as shown
in Figs.~\ref{fig:ds_n}, \ref{fig:ts_n}, and \ref{fig:qs_n} for $U=10$.
All this shows that the ground state of the $1/n$-filled SU($n$) Hubbard 
model is spatially homogeneous, the translational symmetry is not broken.

\subsubsection{Models with $q>n$}

We have chosen as an example $n=3$ and $f=2/5$. As seen in Fig.~\ref{fig:s_l_fig2}(a), 
the block entropy $s_N(l)$ oscillates with period $5$. When every fifth 
data points are fitted to (\ref{eq:cardy}), $c=3$ is obtained. This indicates that the model 
remains critical for finite $U$ as well.  The finite peak in $|\tilde s(k)|$ at
$k^*=4\pi/5$ is due to soft modes. The amplitude of the peak 
disappears in the $N \rightarrow \infty$ limit, the ground state of the system 
is uniform. This is confirmed by the calculation of $\tilde s(k^*)$, $n(k^*)$, 
$D_s$, $T_s$, and $Q_s$ shown in Figs.~\ref{fig:sq_n}, \ref{fig:nq_n}, 
\ref{fig:ds_n}, \ref{fig:ts_n}, and \ref{fig:qs_n}, respectively. There is neither 
a site- nor a bond-centered oscillation in the occupation number or 
bond strength. 


\subsubsection{Models with $q<n$}

\begin{figure}[htb]
\includegraphics[scale=0.5]{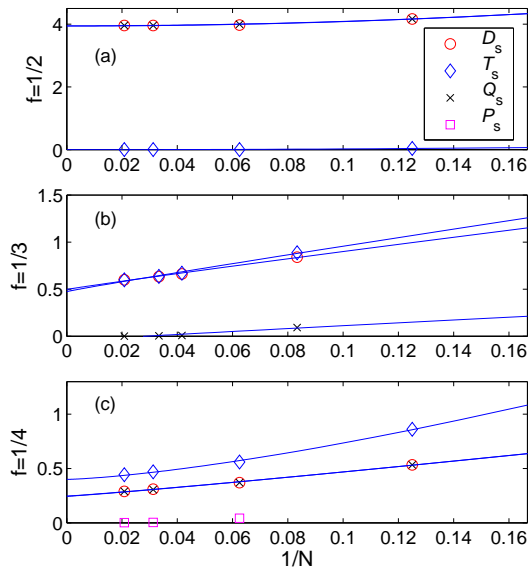}
\caption{Finite-size dependence of $D_s, T_s$, $Q_s$, and $P_s$ for the
half-, one-third-, and quarter-filled one-dimensional SU(5) Hubbard 
model at $U=10$.}
\label{fig:su5_ds}
\end{figure}

One realization of this condition, the half-filled case for $n>2$ has been
studied by us earlier.\cite{buchta_sun} It was found, as shown in 
Fig.~\ref{fig:s_l_fig1}, that the block entropy oscillates with period $2$ 
for any $n > 2$. The peak  in $|\tilde s(k)|$ at $k^*=\pi$ does not 
vanish in the thermodynamic limit (see Fig.~\ref{fig:sq_n}).  In agreement with 
this $D_s$ and $Q_s$ are finite and converge to the same value as shown 
in Figs.~\ref{fig:ds_n} and \ref{fig:qs_n}, while $T_{\rm s}$ vanishes (see 
Fig.~\ref{fig:ts_n}). The same behavior is found in the $n=5$ model at
half filling, as seen in the upper panel of Fig.~\ref{fig:su5_ds}.

On the other hand, $s_1(k^*)$ and $n(k^*)$ vanish in the thermodynamic limit
 (see Fig.~\ref{fig:nq_n}). The translational symmetry of the Hamiltonian is broken, 
the ground state of the half-filled model is dimerized. Stronger and 
weaker bonds alternate along the chain. 

In the one-third-filled case of the $n=4$ and $n=5$ models ($q<n$) $s_N(l)$ 
oscillates with period $3$ and the amplitude of the Fourier component 
$\tilde s(k^*=2\pi/3)$ remains finite even for $N\rightarrow\infty$ (see Figs.~\ref{fig:sq_n} 
and \ref{fig:su5_sq_n}). This indicates that the spatial periodicity 
is tripled in the ground state.  This is corroborated by our results shown in
Figs.~\ref{fig:ds_n}, \ref{fig:ts_n}, \ref{fig:qs_n}, and the middle panel of \ref{fig:su5_ds}.
$D_s$ and $T_s$ scale to the same finite value while $Q_s$, $s_1(k^*)$, 
and $n(k^*)$ vanish. In the ground state, two bonds of equal strength are followed 
by a weaker or stronger bond.

\begin{figure}[htb]
\includegraphics[scale=0.5]{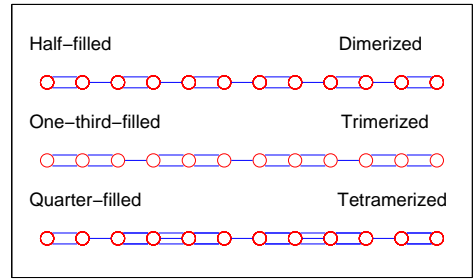}
\caption{Schematic plot of the local bond strength in half-, one-third, and quarter-filled 
SU(5) Hubbard chains.}
\label{fig:su5_bond}
\end{figure}

As a last example we have studied the quarter-filled SU(5) Hubbard chain.
We found that $|\tilde s(k)|$ scales to a finite value at $k^*=\pi/4$ as shown in 
Fig.~\ref{fig:su5_sq_n}. All Fourier components of 
the site entropy and local charge density vanish for long chains, while 
$D_s$, $T_s$, and $Q_s$ scale to finite values. Only $P_s$ scales to zero,
as shown in the lower panel of Fig.~\ref{fig:su5_ds}. The ground state is a 
bond-ordered tetramerized state. Fig.~\ref{fig:su5_bond} shows
schematically the periodic modulation of the bond strength along the
chain for half-, one-third, and quarter-filled models.

\section{Conclusion}

To study the role of multiparticle umklapp processes, we have treated the one-dimensional 
SU$(n)$ Hubbard model analytically by bosonization approach and numerically using the 
DMRG method for $n=3,4$, and $5$ for commensurate fillings $f=p/q$ where $p$ and 
$q$ are relative primes.

Our results confirm that umklapp processes play essentially different role depending on 
the relationship between $q$ and $n$.  When $q=n$ (this is the case in the $1/n$-filled
case) the charge and spin modes are not coupled, the umklapp processes open gap only 
in the spectrum of charge modes. The system remains critical with $n-1$ gapless
spin modes, the central charge is $c=n-1$, and the translational symmetry of the 
Hamiltonian is not broken in the ground state. 

When $q>n$, the leading-order umklapp processes are forbidden in the model
with local interaction by Pauli's exclusion principle. The model is equivalent to an 
$n$-component Luttinger liquid with $c=n$ and the ground state is spatially uniform 
for $U\geq 0$.  

When, however, $q<n$ the charge and spin modes are coupled by the umklapp 
processes and gap opens in the spectrum of all modes. Even more interestingly,
a spatially nonuniform ground state emerges whose periodicity depend on the filling.
Half-filled models develop a dimerized ground state, trimerized state appears 
in one-third-filled models, the ground state is tetramerized in quarter-filled
models. Other periodicities would probably be found at other fillings. Our findings 
are summarized in Table~\ref{tab:sun_numerical} which can be compared to 
the analytical results given in Table~\ref{tab:analitikus}.

\begin{table}[htb]
\begin{tabular}{@{}ccccccccccc@{}}
\hline \hline 
       & \phantom{+} & $n$ & \phantom{+} & $p/q$ & \phantom{+} & $c$ &
      \phantom{+}  & periodicity & \phantom{+} & $k^*$ \\  \hline  
 $q=n$ &  & 2 & & 1/2   & & 1  &  & unform & & $\pi$ \\ 
            &  & 3 & & 1/3   & & 2  &  & uniform & &$2\pi/3$ \\       
            &  & 4 & & 1/4   & & 3  &  & uniform & &$\pi/2$ \\       
            &  & 5 & & 1/5   & & 4  &  & uniform & &$2\pi/5$ \\       
\hline  
 $q<n$ &  & 3 & & 1/2   &  & -  &  & dimerized & & $\pi$  \\       
            &  & 4 & & 1/2   &  & -  &  & dimerized & & $\pi$  \\      
            &  & 4 & & 1/3   &  & -  &  & trimerized & & $2\pi/3$  \\       
            &  & 5 & & 1/2   &  & -  &  & dimerized & & $\pi$  \\       
            &  & 5 & & 1/3   &  & -  &  & trimerized & & $2\pi/3$  \\       
            &  & 5 & & 1/4   &  & -  &  & tetramerized & & $\pi/2$  \\       
\hline  
 $q>n$ &  & 3 & & 2/5  &  & 3  &  & uniform & & $4\pi/5$  \\       
\hline \hline 
\end{tabular}
\caption{Central charge and spatial inhomogeneity for the $p/q$-filled SU($n$) 
Hubbard chain. $k^{\ast}$ in the last column gives the wave number of soft modes 
when the model is critical while it gives the wave number of the nonuniform 
ground state when the model is gapped.}
\label{tab:sun_numerical}
\end{table}

We emphasize that our calculations were performed at a relatively large value of $U$,
where the nonuniformity of the ground state is well developed, and the finite value of the
dimer, trimer or tetramer order parameter can easily be detected. We conjecture, based 
on our earlier calculations,\cite{buchta_sun} that the critical value $U_c$ above which
the nonuniform phase appears, is $U_c=0$. 

\acknowledgments

This research was supported in part by the Hungarian Research Fund (OTKA)
Grants No.\ K 68340, F 046356 and NF 61726 and the J\'anos Bolyai Research
Fund.  The authors acknowledge computational support from Dynaflex Ltd. under
Grant No. IgB-32.

\end{document}